\documentclass[useAMS,usenatbib,usegraphicx]{mn2e}

\usepackage{amssymb}
\usepackage{natbib}
\usepackage{url}

\def \sb {mag\,arcsec$^{-2}$}
\def \muo {$\mu_{0}$}
\def \ks {Kolmogorov-Smirnov}
\def \msun {M$_{\odot}$}
\def \mstar {$M_{\star}$}

\title[Secular evolution of discs in massive barred galaxies]{Evidence for Secular Evolution of Disc Structural Parameters in Massive Barred Galaxies}

\author[S\'anchez-Janssen \& Gadotti]{Rub\'en S\'anchez-Janssen\,$^{1}$\thanks{E-mail:
rsanchez@eso.org} and Dimitri A. Gadotti\,$^{1}$\\
$^{1}$European Southern Observatory, Alonso de C\'ordova 3107, Vitacura, Santiago, Chile\\}

\begin{document}

\date{2013 March 12}

\pagerange{\pageref{firstpage}--\pageref{lastpage}} \pubyear{2012}

\maketitle

\label{firstpage}

%%%%%%%%%%%%%%%%%%%%%%%%%
\begin{abstract}
We address the effects of bar-driven secular evolution in discs by comparing their properties in a sample of nearly 700 unbarred and barred (42 $\pm$ 3 per cent of the population) massive  disc galaxies (\mstar\ $\ge 10^{10}$ \msun).
We make use of accurate structural parameters derived from $i$-band bulge/disc/bar decompositions to show that, as a population, barred discs tend to have fainter central surface brightness ($\Delta$\muo\ $\approx 0.25$ mag), and disc scale lengths that are $\approx$\,15 per cent larger than those of unbarred galaxies of the same stellar mass.
The corresponding distributions of \muo\ and $h$ are statistically inconsistent at the $5.2\,\sigma$ and $3.8\,\sigma$ levels, respectively.
Bars rarely occur in high-surface brightness discs, with less than 5 per cent of the barred population having \muo\ $< 19.5$ \sb\ -- compared to 20 per cent for unbarred galaxies. 
They tend to reside in moderately blue discs, with a bar fraction that peaks at $(g-i)_{disc} \approx 0.95$ mag and mildly declines for both bluer and redder colours.
These results demonstrate noticeable structural differences between the discs of barred and unbarred galaxies, which we argue are the result of bar-driven evolution -- in qualitative agreement with longstanding theoretical expectations.
\end{abstract}
%%%%%%%%%%%%%%%%%%%%%%%%%


%%%%%%%%%%%%%%%%%%%%%%%%%
\section{Introduction}
\label{sect:intro}

Secular processes are expected to have played a significant role in establishing the current properties of massive disc galaxies. These are dynamically evolved systems, as evidenced by the presence of a population of discs at $z\sim1$ with similar scale lengths \citep{Lilly1998,Simard1999,Barden2005} and bar fraction \citep{Jogee2004,Sheth2008} to what is found in the Local Universe. This suggests that the quiescent phase of \emph{massive} discs evolution started at least 8 Gyr ago, thus leaving ample time for secular mechanisms to operate (but see \citealt{Hammer2005}).

Bars perhaps provide the most clear-cut evidence of the impact of secular evolution on disc galaxies. Analytical and numerical calculations show that they drive a significant amount of mass and angular momentum redistribution in the disc \citep{Hohl1971,Sellwood1993,Athanassoula2003,Martinez-Valpuesta2006}, funneling material towards the galaxy inner regions that can result in enhanced gas and stellar densities -- thus possibly fueling nuclear starbursts and AGNs \citep{Shlosman1989} and leading to the formation of central bulge-like components \citep{Courteau1996,Kormendy2004,Debattista2006}. Furthermore, the angular momentum exchange between the inner and outer disc can lead to increased disc scale lengths, and the development of mass profile breaks at large radii \citep{Valenzuela2003,Debattista2006}.  It is precisely the amount of angular momentum exchanged within the galaxy what determines the bar strength, and this ultimately depends on the mass and velocity distributions of the material in the disc and spheroidal (bulge plus halo) components \citep{Athanassoula2003}. Once formed, these non-axisymmetric perturbations appear to be long-lived, surviving buckling instabilities and the growth of a significant central mass concentration ($>$\,10\%, and perhaps up to 20\% of the original disc mass; \citealt{Shen2004,Athanassoula2005,Debattista2006}). On the other hand, \citet{Bournaud2005} suggest that a substantial gas component in the disc ($\sim$\,7\% of the total visible mass) might lead to the weakening, or even destruction of the bar due to the angular momentum exchange resulting from gas inflow.
However, this effect is significantly reduced when a responsive dark matter halo model is considered \citep{Berentzen2007}.

The importance of bars in galaxy evolution studies does not only stem from the fact that they can significantly impact the evolution of a given galaxy, but also because such structures are rather common in disc galaxies. 
%Visual classification using photographic plates yields a bar fraction of 65 per cent in the RC3 catalogue \citep{deVaucouleurs1991}, and 
The most recent optical studies indicate that approximately half of all massive disc galaxies contain bars \citep{Barazza2008,Aguerri2009}. 
The fraction of \emph{strong} bars increases substantially in dust-penetrating near-infrared wavelengths ($\approx$\,60 per cent, \citealt{Eskridge2000}), while the \emph{total} bar fraction does so by up to 35 per cent, either when bar detection is carried out through visual classification \citep{Buta2010} or via quantitative methods \citep{Marinova2007,Menendez-Delmestre2007,Weinzirl2009}. 
It is now well established, though, that the bar fraction in the Local Universe is a strong function of the galaxy stellar mass (\citealt{Mendez-Abreu2010,Mendez-Abreu2012,Nair2010,Cameron2010}), and thus consistent results can only be obtained when comparing samples well matched in \mstar.
Even though wavelength coverage, detection technique, and --possibly to a larger extent-- sample selection, all play a role in the detection of bars in disc galaxies, most studies point to fractions $\sim$\,50 per cent in $L \gtrsim L^{*}$ galaxies.

The picture that emerges from theoretical work is corroborated, at least qualitatively, by a number of observational results. The bar-driven redistribution of angular momentum affects the interstellar medium in galaxies, resulting in flatter chemical abundance gradients in barred galaxies \citep{Martin1994,Zaritsky1994}, as well as higher central concentrations of molecular gas \citep{Sakamoto1999,Sheth2005}. More recently, evidence has been found that the gas brought to the centre by bars is efficiently transformed into stars. \citet{Ellison2011} present indication that the current star formation rate at the centre is higher in massive barred galaxies. \citet{Coelho2011} show that the distribution of mean stellar ages in bulges of massive barred galaxies shows a peak at low ages that is absent for their unbarred counterparts \citep[see also][]{Perez2011}.

Yet investigation of the effects of bar-driven secular evolution on the structural properties of discs is currently lacking in the literature. In this Letter we fill this gap by comparing the disc properties in sample of nearly 700 barred and unbarred galaxies. 
%%%%%%%%%%%%%%%%%%%%%%%%%

%%%%%%%%%%%%%%%%%%%%%%%%%
\section{Galaxy Sample}
\label{sect:sample}

\begin{figure*}
\begin{center}
\includegraphics[width=1\textwidth,clip=true]{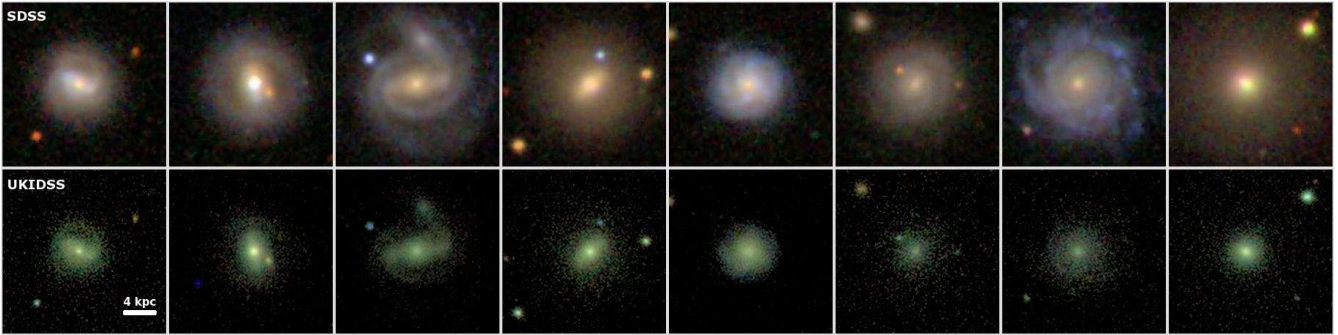}
\end{center}
\caption{SDSS $gri$ (top row) and UKIDSS $YHK$ (botton row) postage-stamp images for a subsample of barred (first four columns) and unbarred (last four columns) disc galaxies of different morphological types in our sample. Bars with semi-major axis lengths $L_{bar} \gtrsim$ 2\,kpc are robustly detected even in the presence of dust (see, e.g., the second barred galaxy).}
\label{fig:pstamps}
\end{figure*}

The sample used here is the one presented in \cite{Gadotti2009a}\,\footnote{See \url{http://www.sc.eso.org/~dgadotti/buddaonsdss.html}}, which contains all galaxies in the SDSS-DR2 with stellar masses \mstar\ $\ge 10^{10}$\,\msun\ \citep[from][]{Kauffmann2003a},\,\footnote{Throughout  this paper we assume a flat cosmology with $\Omega_{M} = 0.3$, $\Omega_{\Lambda} = 0.7$ and $H_{0} = 75$ km\,s$^{-1}$\,Mpc$^{-1}$.}
 at redshift $0.02 \leq z \leq 0.07$, and with axial ratio $b/a\geq 0.9$. These criteria provide a sample which is both representative and suitable for 2D bulge/disc/bar decomposition. The redshift range allows for enough spatial resolution, while selecting face-on galaxies minimizes dust and projection effects, and eases the identification of bars. The reader is referred to that paper for a detailed discussion of selection effects. 

Through multi-band ($gri$) 2D decomposition, \citet{Gadotti2009a} provides accurate structural parameters for the three components (when necessary) -- including the central surface brightness, scale length, integrated colour and stellar mass of the disc component, as well as the galaxy bulge-to-total ratio. 
Despite of the inherent complexity of bulge/disc/bar decompositions, \citet[][Appendix A]{Gadotti2009a} shows that disc structural parameters are particularly stable,  and therefore both $h$ and \muo\ can be robustly derived even in the presence of a bar.
Moreover, it is important to recall that failing to account for the contribution of the bulge results in structural parameters corresponding to a maximum disc, and therefore our multi-component fits provide the least biased results.
%In this Letter we focus on structural parameters derived from fits in the $i$-band, as it is the least affected by dust and recent star formation.
Disc stellar masses were obtained from disc luminosities and mass-to-light ratios in the $i$-band, the latter derived from the integrated $(g-i)$ disc colour and the relation determined by \citet{Kauffmann2007}. 
In order to select a clean sample of disc galaxies, only systems having $B/T<0.8$ have been considered.

To verify whether a galaxy is barred, typical bar signatures were searched for through the inspection of each galaxy image, isophotal contours and a pixel-by-pixel radial intensity profile. 
\citet{Gadotti2011} presents a detailed analysis of the structural properties of the bars in this sample.
It should be noted that, due to the limited spatial resolution of SDSS images, we miss most bars with semi-major axis shorter than $L_{bar} \approx$ 2\,kpc -- which are mainly found in very late-type spirals (later than Sc; \citealt{Elmegreen1985}) and are usually not detected in most recent studies \citep[e.g.,][]{Barazza2008,Aguerri2009}.
At this point it is important to underline once again that the detection of bars and the measurement of structural parameters are carried out in the \textit{i}-band. 
This, together with the low inclination of the sample ($i \lesssim 25^{\circ}$), results in a significant reduction of the effects of dust attenuation.
For instance, the dust extinction models by \citet{Ferrara1999} indicate that at these low inclinations, and for realistic dust optical depths, the typical $i$-band extinction is only $0.05-0.15$ mag larger than in the $K$-band -- while the corresponding values in the commonly used $B$-band are instead $0.2-0.4$ mag larger.
In order to illustrate the robustness of our methodology, and that we are not missing a significant number of bars due to dust or star formation effects, in Fig.\,\ref{fig:pstamps} we show three-colour postage-stamps images in both the optical (SDSS) and NIR (UKIDSS) for four barred and four unbarred galaxies of different morphological types in the sample. Strong bars above our size detection limit are robustly detected even in the presence of dust (see, e.g., the second barred galaxy).
As discussed  above, we are probably missing a good fraction of weak bars, and thus the results from this study concern massive discs hosting strong bars.

Our final sample consists of 291 barred and 393 unbarred disc galaxies, corresponding to a bar fraction of 42 $\pm$ 3 per cent (binomial 90\% uncertainty) -- in excellent agreement with previous studies \citep[e.g.,][]{Aguerri2009}. Both subsamples are perfectly matched in disc stellar mass, with 2.5, 50 and 97.5 per cent quantiles of [0.5,1.7,5.5]$\times10^{10}$ \msun\ and [0.4,1.6,6.5]$\times10^{10}$ \msun\ for barred and unbarred galaxies, respectively.
%%%%%%%%%%%%%%%%%%%%%%%%%

%%%%%%%%%%%%%%%%%%%%%%%%%
\section{Properties of discs in barred and unbarred galaxies}
\label{sect:results}

\begin{figure}
%\begin{center}
\includegraphics[width=0.5\textwidth,clip=true]{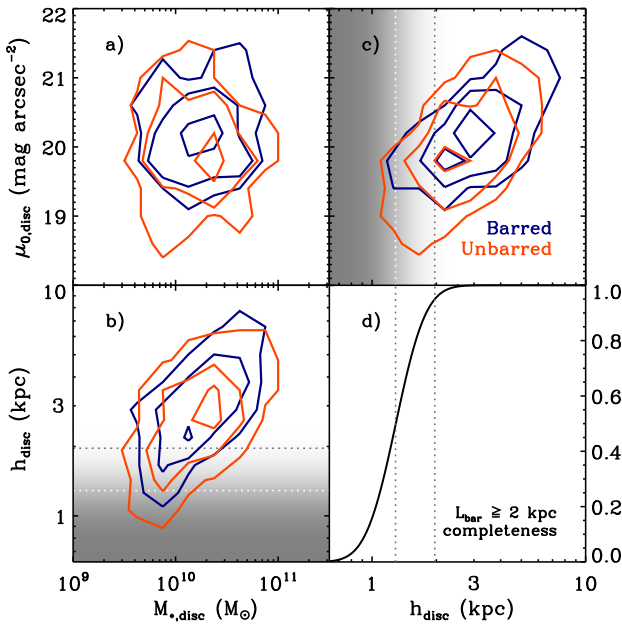}
\caption{
Structural scaling relations of discs in barred and unbarred galaxies. In all panels isocontours enclose 25, 75 and 95 percent of each population. \emph{a)} Disc central surface brightness, \muo, as a function of disc stellar mass. \emph{b)} The disc scale length vs stellar mass relation. \emph{c)} Joint \muo-$h$ relation for both subsamples. \emph{d)} Bar detection function, with dotted lines indicating the 50\% and 95\% completeness values. Shaded regions in the other panels show the same completeness function. All parameters are derived from fits to the $i$-band images. Discs in barred galaxies are characterised by having fainter central surface brightness and larger scale lengths. Note the lack of bars in galaxies having compact, high surface brightness discs.
}
\label{fig:scalings}
%\end{center}
\end{figure}

Figure\,\ref{fig:scalings} reveals the striking differences of the structural scaling relations of galaxy discs in barred and unbarred galaxies. Panels \ref{fig:scalings}$a$ and \ref{fig:scalings}$b$ show, respectively, the $i$-band disc central surface brightness (\muo) and scale length ($h$) as a function of \emph{disc} stellar mass\,\footnote{By using the \emph{current} disc stellar mass instead of the galaxy total stellar mass we eliminate all the possible uncertainties and biases associated to varying bulge-to-disc ratios at fixed \mstar. We also constructed this plot using the bar-plus-disc mass in the case of barred galaxies, to represent the \emph{original} disc mass before the instability developed. This does not modify our results, simply because the bar mass fractions are typically small (median bar-to-total ratio $\sim$\,0.1; \citealt{Gadotti2011}). We however prefer to use the current disc mass as derived from the exponential fits because it is the quantity directly related to $h$ and \muo.}, while panel \ref{fig:scalings}$c$ depicts the bivariate distribution of the two former quantities. Isocontours in all panels enclose 25, 75 and 95 per cent of each population.
There are several remarkable features worth discussing.
First, panel \ref{fig:scalings}$a$ shows that the distribution of disc central surface brightness is significantly different for barred and unbarred galaxies, in the sense that at a given disc stellar mass barred galaxies tend to have fainter \muo\ values -- note how all isocontours extend towards much brighter \muo\  for unbarred galaxies than for barred ones. 
Notably, the reduction in surface brightness is accompanied by an increase of disc scale length (panel \ref{fig:scalings}$b$), resulting in markedly dissimilar distributions in the \muo-$h$ plane (panel \ref{fig:scalings}$c$). While unbarred galaxies populate the high-surface brightness and small-scale lengths parameter space, barred galaxies are essentially absent from this region -- only 5 per cent of all barred galaxies have \muo\ $ < 19.5$ \sb, while this fraction increases up to 20 per cent for unbarred discs. 
One potential explanation for the lack of bars in these discs is that they have semi-major axis lengths $L_{bar} \lesssim$\,2 kpc and thus remain undetected in our analysis. 
In order to quantify this incompleteness, we fit the $L_{bar}-h$ relation presented in \citet{Gadotti2011}. We find $L_{bar}=1.55\,h^{0.9}$, with a 0.1 dex intrinsic scatter. Using this relation and scatter we then compute, for any given $h$, the fraction of discs that may host bars larger than 2 kpc. The resulting bar detection function is shown in Fig.\,\ref{fig:scalings}$d$, and by the shaded areas in the other panels of the same Figure. Dotted lines correspond to the 50\% and 95\% completeness values.
Note that incompleteness effects are negligible for all $h \gtrsim 2$ kpc discs.
From Fig.\,\ref{fig:scalings}$c$ it is clear that the difference in scale length distributions holds for discs in the high completeness region, where the bar detection function exceeds the 50\% value -- and thus shows that the paucity of bars in high surface brightness discs is real and not a selection effect. 
We still detect bars in small discs with $h \lesssim 1.5$ kpc, but they only occur in the lowest surface brightness systems. Additionally, panel $1c$ shows that bars are still missing in large discs having \muo\ $ < 19.5$ \sb, where bar non-detection is not an issue. The remarkable result here is the paucity of long bars in these high surface brightness systems, and the evidence that barred discs extend towards the large-$h$, faint-\muo\ region of the plot.

\begin{figure}
\begin{center}
\includegraphics[width=0.5\textwidth,clip=true]{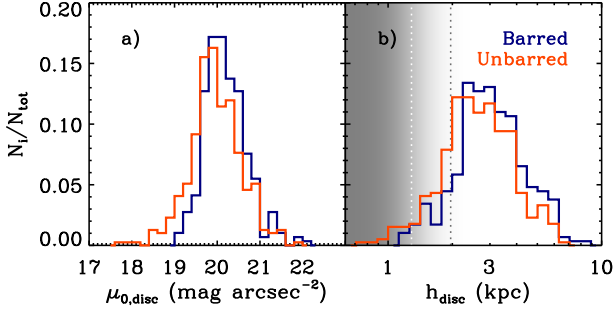}
\caption{
Normalised distributions of disc central surface brightness ($a$) and disc scale lengths ($b$) for barred and unbarred galaxies. The shading represents the bar detection function. According to a \ks\ test, the corresponding distributions of \muo\ and $h$ are statistically inconsistent at the $5.2\,\sigma$ and $3.8\,\sigma$ levels, respectively. As a population, discs in barred galaxies tend to have $\approx$\,0.25 mag fainter central surface brightness and $\approx$\,15 per cent larger disc scale lengths.
}
\label{fig:hmu}
\end{center}
\end{figure}

These differences can be more clearly appreciated in Fig.\,\ref{fig:hmu}, where we show the normalised distributions of \muo\ and $h$ for our two populations (panels $2a$ and $2b$, respectively). The central surface brightness distribution of discs in unbarred galaxies peaks at a moderately more luminous value, and  features a distinct extended tail towards brighter \muo. A \ks\ test rules out the null hypothesis that the two distributions are drawn from the same parent population at the $5.2\,\sigma$ level. The distributions of disc scale lengths for both populations are also inconsistent at the $3.8\,\sigma$ level according to the KS test. The two distributions would however be statistically indistinguishable if barred galaxies had 0.25 mag brighter central surface brightness and 15 per cent smaller scale lengths. It is important to note that, despite the high statistical significance of the differences here reported, there is considerable scatter at a given disc stellar mass: $\sigma_{\mu_{0}} \approx 0.5$ mag and $\sigma_{h} \approx 1.10$ kpc.

\begin{figure}
\begin{center}
\includegraphics[width=0.5\textwidth,clip=true]{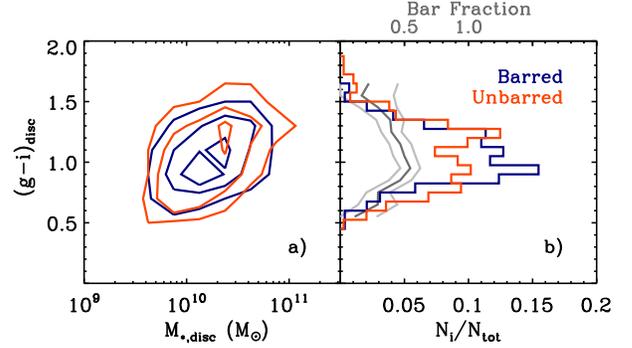}
\caption{
\emph{a)} The $(g-i)_{disc}$ colour vs disc stellar mass relation for barred and unbarred galaxies. \emph{b)} Normalised distributions of disc colours for both subpopulations. The dark gray curve shows the strong dependence of the bar fraction with disc colour, while light gray curves correspond to the binomial 90\% confidence intervals. Bars tend to reside in moderately blue discs, with a bar fraction that peaks at $\approx$\,55\% at $(g-i)_{disc}\approx 0.95$ mag, and declines for both redder and bluer colours.
}
\label{fig:colors}
\end{center}
\end{figure}

Aside from the structural scaling relations, we investigate the stellar population properties of discs in our sample of barred and unbarred galaxies. Fig.\,\ref{fig:colors}$a$ presents the relation between the \emph{disc} stellar mass and its integrated $(g-i)_{disc}$ colour, while panel \ref{fig:colors}$b$ shows the corresponding normalised colour distributions. The disc colour distribution appears to be marginally bimodal for both populations\,\footnote{A one-sided \ks\ test rules out the null hypothesis of a normal distribution at the 90\% (1.6\,$\sigma$) and 99.8\% (3.1\,$\sigma$) levels for barred and unbarred galaxies, respectively.} but, remarkably, their peaks differ from  $(g-i)_{disc} \approx 1.25$ for unbarred galaxies to $(g-i)_{disc} \approx 0.95$ for barred ones. 
The dark gray curve in Fig.\,\ref{fig:colors}$b$ shows the bar fraction as a function of disc colour (only for bins having more than 5 galaxies), while the light gray curves correspond to the binomial 90\% confidence intervals. Bars tend to reside in moderately blue discs, with a fraction that peaks at $\approx$\,55\% at the previously mentioned value and mildly declines for both bluer and redder discs -- in qualitative agreement with previous work \citep{Aguerri2009,Masters2011}.
The fraction of bars in galaxies with very blue colours is probably a lower limit (cf. \citealt{Nair2010}), as we recall that these late-type systems can contain small bars that are undetected in our images.

%%%%%%%%%%%%%%%%%%%%%%%%%

%%%%%%%%%%%%%%%%%%%%%%%%%
\section{Discussion and conclusions}
\label{sect:discussion}

We have shown that barred and unbarred \mstar\ $\ge 10^{10}$ \msun\ galaxies are characterised by distinct distributions of disc structural parameters.
As a population, barred discs tend to have fainter central surface brightness ($\Delta$\muo\ $\approx 0.25$ mag in the $i$-band) and disc scale lengths that are larger by $\approx$\,15 per cent than those of unbarred galaxies.
We argue here that these differences are the result of bar-driven disc evolution. While the alternative scenario --i.e., that the distinction arises due to initial disc conditions that favoured bar formation-- is difficult to rule out solely based on observations, theoretical work strongly suggest that is not the case.
First, all numerical simulations show that the onset of the bar instability results in structural evolution of the disc (to varying degrees). Given that we are studying barred galaxies, it is reasonable to expect that the disc has actually evolved, i.e., that its current properties are not exactly the same ones that led to the formation of the bar. Second, and most important, the alternative scenario would imply that bar formation is favoured in discs having lower surface brightness and larger scale lengths. This is in disagreement with the numerical results by \citet{Mayer2004}, where it is shown that low surface brightness discs are generally rather stable against bar formation due to a combination of low self-gravity and a high halo/disc mass ratio.

The observed differences in disc scale lengths are close to, but slightly lower than,  those predicted by the numerical simulations of \citet{Valenzuela2003}, where discs were found to increase their $h$ by factors 1.2$-$1.5 due to the transfer of angular momentum that accompanies the formation of a bar. 
Nevertheless, the moderate increment of scale lengths  revealed by the data has to be thought of as a lower limit to the actual effects of bar-driven secular evolution. The reason is twofold.
First, it is important to recall that bars are not the only drivers of disc secular evolution.  In fact, any type of non-axisymmetric perturbation --including spiral arms, oval distortions and triaxial dark matter haloes-- can also modify the properties of discs (e.g., \citealt{Kormendy2004,Sellwood2010}), but they generally operate on longer timescales. It is therefore possible that secular evolution has also occurred, to some degree, in our sample of unbarred galaxies.
Second, our disc structural parameters come from fitting one single exponential to the surface brightness profile, with no distinction between different disc Freeman types \citep{Freeman1970}. This can slightly bias our recovered parameters, resulting in marginally brighter \muo\ and smaller $h$ in the case of Type\,II profiles with outer truncations -- which occur far more often than single exponential, Type\,I profiles \citep[e.g.,][]{Pohlen2006}.
This is indeed supported by the recent analysis of disc properties in S$^{4}$G galaxies \citep{Sheth2010}. Mu\~noz-Mateos et al. (in preparation)  find that, when allowing for outer profile breaks, disc scale lengths in barred galaxies are, on average, a factor $\sim$\,1.8 larger than those without bars -- but, as in our case, the scatter at fixed 3.6\,$\mu$m magnitude is large.

In any case, our fits provide the optimal average exponential profile between small and large radii, and thus allow for a direct comparison with the numerical simulations by \citet{Debattista2006}. 
Not surprisingly, they find that the amount of evolution in their simulated discs depends critically on the \emph{initial} disc kinematics. The density profiles of highly unstable discs (low Toomre's $Q$) evolve dramatically compared to more stable initial configurations, resulting in final scale length differences of factors $\gtrsim$\,2 even for models with nearly identical initial angular momentum. 
The general situation is of course more complex than this, and the specific density profile evolution in their simulations is determined by the phase-space distributions of the stellar disc and  the dark matter halo -- resulting in scale lengths changing by factors 1.0$-$2.4. 
As \citet{Debattista2004,Debattista2006} point out, this has the important consequence that direct estimates of dark matter halo spin parameters from measured disc scale lengths can be rather uncertain.

Our results support their claims, but suggest that the amount of scale length evolution due to bar formation is only moderate.
A simple back-of-the-envelope argument appears to be consistent with this idea. If secular mechanisms were to increase disc scale lengths by a considerable amount --say, factors larger than two--, we would then expect to see clear signs of  evolution in the mass-size relation of discs between redshifts $0 < z < 1$.
Yet observations support mild to no evolution at all \citep{Lilly1998,Simard1999,Barden2005} and this, in turn, is consistent with real discs being reasonably stable ($Q \gtrsim 1.5$; e.g., \citealt{Kregel2005}).
If this is the case, and considering all the assumptions involved, it is most likely that the uncertainties derived from mapping halo spin to disc scale lengths are not dominated by secular evolution effects -- but they certainly are a contributing factor.

We note here that we have also looked for any existing correlation between the bar strength, as measured by its ellipticity, and the central surface brightness and scale lengths of the discs. No trend whatsoever is found between $\epsilon_{bar}$ and $\mu_{0}$. While the mean bar ellipticity changes from $\langle \epsilon_{bar} \rangle \approx 0.6$ at $h\sim2$ kpc to $\langle \epsilon_{bar} \rangle \approx 0.7$ at $h\sim5$ kpc, the 0.1 scatter at fixed $h$ renders any potential trend statistically insignificant (a Spearman coefficient of only 0.3).
The lack of significant correlations between these quantities suggests that the impact of bars on galaxy discs is rather complex, and does not solely depend on bar strength. This is in line with theoretical work indicating that a number of factors play an important role in the formation and evolution of bars.

Indeed, it is not yet clear if there exists one single condition determining whether a galaxy will be bar-stable or not.
\citet{Athanassoula2008} shows that simple criteria that do not fully capture the complexity underlying bar formation [e.g., the \citet{Efstathiou1982} criterion] generically fail to  correctly predict  disc stability.
Instead, numerous factors come into scene in order to stabilise a disc against non-axisymmetric perturbations. 
Thus, galaxies with very weak or no bars must have either a kinematically hot disc and/or a significant central mass concentration and/or a very low relative disc mass and/or be embedded in a quite unresponsive dark matter halo \citep{Athanassoula2003,Sellwood2010}.
From the disc component point of view, a high stellar velocity dispersion can provide significant stabilisation and prevent the formation of a bar for over a Hubble time \citep{Athanassoula1986}.
Moreover, \citet{Sellwood2001} show that a steeply rising inner rotation curve is sufficient to bar-stabilise a disc, regardless of the dark matter content.
In this context, the realisation in Fig.\,\ref{fig:scalings}$c$ that compact, high-surface brightness discs do \emph{not} host bars --or if they do, they are rather small-- is intriguing, and it is tempting to think of these systems as having a high rate of shear at the centre capable of (almost) fully stabilising the disc. Kinematical studies of these galaxies are highly desirable in order to test this hypothesis and clarify their lack of bars.

Finally, basic stability criteria and numerical simulations suggest that bars are more easily triggered in gas-rich (i.e., cold), star-forming discs. This provides a natural explanation for the preferential occurrence of bars in moderately blue discs, but the exact shape of the bar fraction distribution deserves further investigation.
Even though such an analysis is beyond the scope of this Letter, one can naively think that gas consumption through star formation leads to redder and kinematically hotter
(more bar-stable) discs. In this context, the declining fraction towards bluer colours has probably to be understood in terms of a selection effect, such that these galaxies contain smaller bars that remain undetected in our images (cf. \citealt{Nair2010}).
 
We have shown that  the discs of barred and unbarred galaxies are characterised by distinct structural properties, which we argue are the result of bar-driven evolution.
%bars drive noticeable structural evolution of galaxy discs, confirming longstanding analytical and numerical predictions.
Detailed structural decomposition of galaxies provides one of the most powerful diagnostics for galaxy evolution studies, allowing for direct, quantitative comparisons with theory and  simulations.

%%%%%%%%%%%%%%%%%%%%%%%%%

\section*{Acknowledgements}
We acknowledge an anonymous referee for suggestions that improved the clarity of the manuscript and the presentation of results. We thank J. M\'endez-Abreu, I. P\'erez, J. Falc\'on-Barroso, I. Trujillo and S. Courteau for useful comments and discussions. Funding for this research was provided in part by the Marie Curie Actions of the European Commission (FP7-COFUND). This work is based in part on data obtained as part of the Sloan Digital Sky Survey and the UKIRT Infrared Deep Sky Survey.

\bibliographystyle{mn2e}
\bibliography{/Users/rsanchez/WORK/PAPERS/rsj_references.bib}
%\begin{thebibliography}{99}
%\end{thebibliography}

\label{lastpage}

\end{document}